\documentclass[12pt]{article}
\usepackage{amssymb}
\usepackage{amsmath, amsthm}
\newcommand{\be}{\begin{equation}}
\newcommand{\ee}{\end{equation}}
\begin{document}
\def\theequation{\arabic{section}.\arabic{equation}}
\begin{titlepage}
\title{Non-chaotic dynamics in general-relativistic 
and scalar-tensor cosmology}
\author{V. Faraoni, M.N. Jensen and S.A. Theuerkauf \\ \\
{\small \it Physics Department, Bishop's University}\\
{\small \it Lennoxville, Qu\`{e}bec, Canada J1M 1Z7}
}
\date{} \maketitle
\thispagestyle{empty}
\vspace*{1truecm}
\begin{abstract}
In the context of scalar-tensor models of dark energy and 
inflation, the dynamics of vacuum 
scalar-tensor  cosmology are analysed without specifying the 
coupling  function or the scalar field potential. A conformal 
transformation to the Einstein frame is used and the 
dynamics of general relativity with a minimally coupled scalar 
field are derived for a generic  potential. It is  shown that 
the 
dynamics are 
non-chaotic, thus settling an existing debate.
\end{abstract} \vspace*{1truecm}

\begin{center}  {\bf Keywords:} scalar-tensor cosmology, dark 
energy, inflation.
\end{center}
\begin{center} {\bf PACS:}   98.80.-k, 04.90.+e, 04.50.+h
\end{center}
\setcounter{page}{0}
\end{titlepage}

\def\theequation{\arabic{section}.\arabic{equation}}


\section{Introduction}
\setcounter{equation}{0}
 \setcounter{page}{1}
Recent cosmological observations have established that the
universe is very close to being spatially flat, corresponding to
vanishing curvature index $ K $ in the
Friedmann-Lemaitre-Robertson-Walker (``FLRW") line element 
\be
\label{1} 
ds^2=-dt^2+a^2(t)\left[ \frac{dr^2}{1-Kr^2} +r^2\left(
d\theta^2 + \sin^2 \theta \, d\varphi^2 \right) \right] 
\ee 
in comoving
coordinates $\left( t,r, \theta ,\varphi \right)$. We use units 
in which the speed of light and Newton's constant  assume the 
value unity, the metric signature is
$-,+,+,+ $, and the other notations follow those of
Ref.~\cite{Wald}. If $ K $ 
were exactly zero, it would not be possible to establish it 
observationally, due to
unavoidable experimental errors. Notwithstanding this, the 
cosmic microwave background experiments \cite{CMB} measuring a 
total energy density $\rho$ of the universe
close to the critical density $\rho_c=\frac{3H^2}{8\pi G}$ 
(where
$H \equiv \dot {a}/ a $ is the Hubble parameter and an overdot
denotes differentiation with respect to the comoving time $t$) 
can be
regarded as a reassuring validation of the theoretical prediction
that the universe was taken extremely close to the $ K =0$ state
by inflation early in its history \cite{LiddleLyth}.

Another surprising discovery, obtained with the study of 
type Ia
supernovae at high redshifts \cite{SN}, is that the cosmic 
expansion
is accelerated. In the context of Einstein's gravity this fact is
explained by postulating that the pressure $P$ of the cosmic 
fluid
satisfies $P < -\rho/3 $. In fact, the best fit 
to the
observational data requires even more exotic properties for the
{\em dark energy}, the fluid that accounts for 70\% of the
energy density of the universe in Einstein's theory; if $w\equiv
P/\rho $ denotes the effective equation of state parameter of
the dark energy, there is marginal evidence 
that $w < -1 $ \cite{w}. This range of values of the 
parameter $w$ 
corresponds to a Hubble parameter that is increasing with 
time ({\em superacceleration}):
$\dot{H} > 0$  according to the Friedmann equation
\be \label{2222} 
\dot{H}= - \, \frac{\kappa}{2}(P+\rho) \;,
\ee
where $\kappa \equiv  8\pi G $ and $G$ is Newton's constant. 
Most dark energy models are 
based on scalar fields and, if the universe really 
superaccelerates, models based on general relativity with a 
canonical, minimally coupled, scalar field $\phi$ are unviable. 
In fact, the energy density and pressure of such a scalar field 
are
\be\label{3} 
\rho = \frac{\dot{\phi}^2}{2}+ V(\phi) \;, \;\;\;\hspace{10 mm} 
P =\frac{\dot{\phi}^2}{2} -V(\phi) \;, 
\ee
where $V(\phi)$ is the scalar field potential, and 
eq.~(\ref{2222}) 
reduces to $\dot{H} =  -\kappa \dot{\phi}^2 /2 \le 0 $ 
(the limiting situation $\dot{H} =0$ describes de Sitter solutions).
Furthermore, the best fit to the observational data requires a
dynamical form of dark energy with an equation of state 
parameter
changing with redshift, which would definitely exclude the
cosmological constant as an explanation for dark energy (such a 
model
is anyway unappealing because of the cosmological constant problem
and the cosmic coincidence problem 
\cite{Weinberg1,Weinberg2}).  For this reason, alternative 
models in which the universe can
superaccelerate have been considered, including phantom fields with
negative kinetic energy \cite{phantom}, scalar fields coupled
non-minimally to the curvature \cite{NMC,superquintessence}, or
alternative gravity theories. In this paper, we consider
scalar-tensor extensions of general relativity. The theories in 
this class \cite{ST,Will} generalize  Brans and Dicke's 
\cite{BransDicke} theory, exhibit many features in common with 
string/M theory \cite{string}, are the arena for extended 
\cite{extended} and hyperextended 
\cite{hyperextended} inflationary scenarios, and are 
widely used in cosmology \cite{mybook,FujiiMaeda}. 
Scalar-tensor gravity is used to model
inflation in the early universe or dark energy in the present
late time era of the universe. Both phantom cosmology and the 
theory of a non-minimally
coupled scalar field, capable of producing the superacceleration
phenomenon, can be seen as scalar-tensor theories.

Because of the wide use of scalar-tensor gravity, motivated by 
the belief that the latter may be more fundamental than 
Einstein's gravity,
one would like to have a clear and general picture of scalar-tensor
cosmology. In general relativity the dynamics of a particular (dark
energy or inflationary) scenario is usually determined once the
potential is fixed; in scalar-tensor gravity one also has  
free 
coupling functions adding extra degrees of freedom. However, it
would be desirable to understand the dynamics with as much 
generality  as possible without fixing these functions. This is 
what we set out to do in
this paper and, on the basis of the observational data and of
theoretical prejudice, we restrict ourselves to considering a
spatially flat ($ K=0$) universe. In particular, we derive
conclusions about the dynamics based on general assumptions 
about the
potential (e.g., monotonic, {\em etc.}). To the best of our 
knowledge, this classification has not been explored even in the 
relatively simple case
of general relativity with a minimally coupled scalar field, in
which the potential $V(\phi)$ is the only unknown function 
(e.g., in inflationary scenarios). Moreover, the geometry of the 
phase space is rarely discussed, even in scenarios based on 
general relativity with
specific choices of the potential $V(\phi)$, for which phase space
studies exist in the literature (see 
Refs.~\cite{KolitchEardley}-\cite{HoldenWands} 
for Brans-Dicke theory) --- 
usually only projections of the phase space are considered. 
The role of chaos in cosmology has received much attention since 
the early work on anisotropic universes approaching an initial 
singularity  (mixmaster universes) \cite{mixmaster}.

A particularly interesting aspect of the cosmological dynamics is
the presence or absence of chaos in a $ K =0 $ FLRW universe 
with a single scalar field; this has been the subject of debate. 
While
the presence of chaos in the dynamics has been suggested on the
basis of numerical studies and of a Painlev\'e analysis 
\cite{Blancoetal,HelmiVucetich}, it has been stated explicitly 
that chaos can not occur in the case of a single, minimally or 
non-minimally coupled, scalar field \cite{GunzigetalMPLA}. This 
has led to regarding
inflation in the early universe as playing a new role: by taking the
universe extremely close to a spatially flat one, primordial
inflation eliminates the possibility of chaos from the
dynamics.  This statement is based on an analytic study of the 
phase space and its dimensional reduction to a two-dimensional 
surface \cite{GunzigetalMPLA,GunzigetalCQG}. Since this 
dimensional reduction can be generalized to 
any scalar-tensor $ K =0$ cosmology with a  single scalar
field, it would appear 
that chaos is absent. 
However,
the statement is based on the two-dimensional nature of the phase
space but the known theorems (e.g., \cite{Glendinning}) do not apply
to a multi-sheeted, non-compact and non-connected phase space 
such as
the one of scalar-tensor cosmology. Therefore, the problem of the
presence or absence of chaos remains open. We show in 
the following that, while determining the dynamics for general
potentials, one is also able to exclude the presence of chaos in 
the dynamics of scalar-tensor $ K=0 $ cosmologies.

In section~2 we approach the problem of determining the 
dynamics in the context of (spatially flat) scalar-tensor 
cosmology in the Jordan frame, while in section~3 we reduce it 
to the equivalent problem in the context of Einstein's theory by 
means of a conformal transformation to the Einstein frame. In 
section~4 we discuss the dynamics of the scale factor and of the 
scalar field for general potentials in general relativity, and 
then the conformal transformation is used to map back the 
results to the Jordan frame. Section~5 contains a discussion and 
the conclusions.


\section{Scalar-tensor cosmology }
\setcounter{equation}{0}

We consider the scalar-tensor class of
theories \cite{ST} described by the general action
\be \label{4} 
S=\frac{1}{16\pi}\int d^4 x \sqrt{-g} \left[\phi R -
\frac{\omega(\phi)}{\phi} g^{ab}\nabla_a \phi \nabla_b \phi -
V(\phi) \right] \;.
\ee
We do not add to the action terms describing ``ordinary" 
matter (as 
opposed to  scalar field $\phi$) because we are 
interested in situations in which the cosmic dynamics are 
dominated by the scalar field, e.g.,  during early inflation or 
during the late time dark energy era. The field equations are
\begin{eqnarray}
 R_{ab}-\, \frac{1}{2} \, g_{ab} R &=&  
\frac{\omega(\phi)}{\phi^2} 
\left(\nabla_a\phi\nabla_b\phi-\frac{1}{2} \,
g_{ab} \nabla^c\phi\nabla_c\phi \right) \nonumber \\
&&\nonumber \\
&+& 
\frac{1}{\phi}\left(\nabla_a \nabla_b \phi - g_{ab}\Box
\phi\right)-\frac{V(\phi)}{2\phi}g_{ab} \;, \label{5}\\
&& \nonumber\\
  \Box\phi &=& \frac{1}{2\omega(\phi)+3} \left[\phi \, 
\frac{dV}{d\phi} - 2V(\phi) -\frac{d\omega}{d\phi} \, \nabla^c 
\phi
\nabla_c \phi \right] \;,  \label{5bis}
\end{eqnarray}
where $\nabla_a$ is the covariant derivative operator associated 
with the metric $g_{ab}$, and $\Box\equiv g^{ab} \nabla_a 
\nabla_b$. In a $ K=0$ FLRW metric 
these equations assume the 
form
\begin{eqnarray}
&& H^2 = -H\, \frac{\dot{\phi}}{\phi}+\frac{\omega(\phi)}{6}
\left( \frac{\dot{\phi}}{\phi}\right)^2 + \frac{V(\phi)}{6\phi} 
\;, \label{7} \\
&&\nonumber\\
&& \dot{H} = -\, \frac{\omega(\phi)}{2}
\left(\frac{\dot{\phi}}{\phi}\right)^2 +
2H\left(\frac{\dot{\phi}}{\phi}\right)  
+\frac{1}{2\phi \left( 2\omega(\phi)+3 \right) }\left[ \phi V'- 
2V+  \omega' \left( \dot{\phi} \right)^2 \right] \;,  \label{8} 
\\
&&\nonumber\\
&& \ddot{\phi} + \left( 3H + \frac{ \dot{\omega} }{2\omega+3} 
\right) \dot{\phi}= \frac{1}{2\omega+3}\left( 2V-\phi V' \right) 
\;,  \label{9} 
\end{eqnarray}
where a prime denotes differentiation with respect to $\phi$. 
Eqs.~(\ref{7})-(\ref{9})  are respectively, the Hamiltonian 
constraint $H^2 = \kappa\rho_\phi/3 $, $\dot{H} = 
-\kappa \left(\rho_\phi+P_\phi\right)/2 $, and the Klein-Gordon 
equation for $\phi$, where
$\rho_\phi$ and $P_\phi$ are the effective energy density and
pressure of the scalar. Only two equations in the set 
(\ref{7})-(\ref{9}) are independent. We choose $H$ and $\phi$ 
as dynamical
variables (other works in the literature choose different
variables for ease of manipulation in special cases, but the 
physical meaning of
their results is somewhat obscured). To analyze the phase space it
is convenient to regard the Hamiltonian constraint (\ref{7}) as the
algebraic equation for $\dot{\phi}$
\begin{equation}
\label{10} \omega \dot{\phi}^2 -6H\phi\dot{\phi} + \left(\phi V -
6H^2\phi^2\right) =0
\end{equation}
with roots
\begin{equation}
\label{11} \dot{\phi}_\pm \left(H,\phi\right) =
\frac{1}{\omega(\phi)} \left(3H\phi \pm
\sqrt{3H^2\phi^2\left(2\omega+3\right)-\omega\phi V } \, 
\right).
\end{equation}
Unless $\omega=0$, the phase space is a two-dimensional 
surface $\Sigma = \Sigma^+ 
\cup
\Sigma^-$ composed of two sheets $\Sigma^\pm$ corresponding to the
positive or negative sign in eq.~(\ref{11}), embedded in the
three-dimensional space $(H,\phi,\dot{\phi})$. In actual fact, 
only the region $\phi > 0$ is of physical interest because it 
corresponds to
positive gravitational coupling, and there are physical constraints
on the variability of the effective gravitational coupling
\begin{equation}
G_{eff}(\phi) =
\frac{2\left(\omega+2\right)}{\left(2\omega+3\right)\phi}
\end{equation}
(see, e.g., Refs.~\cite{mybook,FujiiMaeda}).

In general, there is a dynamically forbidden region 
\begin{equation}
\label{12} {\cal F} = \left\{ \left(H,\phi\right): 
\;\;\; 3H^2\phi^2
\left[2\omega(\phi)+3\right] - \omega(\phi)\phi V(\phi) <0 \right\},
\end{equation}
corresponding to a negative argument of the square root in
eq.~(\ref{11}). This forbidden region may be absent for 
specific
choices of $\omega(\phi)$ and $V(\phi)$, but is present in
most scenarios considered in the literature (see Ref. 
\cite{FaraoniAnnPhys} for further details). The boundary of the
forbidden region is composed of the only points where the two sheets
touch each other and $\dot{\phi}$ is single valued, i.e., the set
\begin{equation}
\label{13} 
{\cal B} \equiv \left\{\left( H,\phi \right): \;\;\;\dot{\phi} =
\frac{3H\phi}{\omega(\phi)} \, \right\} = \Sigma^+\cap\Sigma^- 
\;,
\end{equation}
or
\begin{equation}  \label{14}
\left( H,\phi \right):\;\;\;  3H^2\phi^2\left( 2\omega+3\right) 
= \phi \, \omega \, V
\end{equation}
on ${\cal B}$, which defines implicitly a curve $H(\phi)$ in the
space $(H,\phi,\dot{\phi})$.

Having chosen $H$ and $\phi$ as dynamical
variables, the equilibrium points of the system are necessarily de
Sitter spaces with constant scalar field $(H_0,\phi_0)$. If they
exist, these can lie anywhere in the phase space 
$\Sigma \cap \left\{ \dot{\phi}=0 \right\}$.   According
to the dynamical equations (\ref{7})-(\ref{9}), there are two
necessary and sufficient conditions for the existence of such de
Sitter fixed points.
\begin{eqnarray}
&&  H_0^2 = \frac{V(\phi_0)}{6\phi_0} \;, \label{14bis}  \\
&& \nonumber \\
&&\phi_0 \, V_0'-2V_0 = 0 \;.  \label{14ter} 
\end{eqnarray}
This two-dimensional structure of the phase space was reported
earlier in the literature for minimally \cite{Foster} and
non-minimally coupled  \cite{ALO,GunzigetalCQG} scalar fields. In 
Ref.~\cite{GunzigetalMPLA} it was argued that the reduction of 
the phase space to two dimensions implies that chaos is 
impossible, and therefore that
primordial inflation taking the universe extremely close to an
exactly spatially flat FLRW space in addition to solving the
problems of standard Big Bang cosmology \cite{LiddleLyth}, has 
the effect of
inhibiting chaos. However, this conclusion is based on the
Poincar\`{e}-Bendixson theory which applies to a 
two-dimensional
phase space that is flat instead of curved, and to regions that are
compact and connected. This is certainly not the case of the 
$\Sigma$ phase
space which, in addiction to being double-sheeted and curved,
extends to infinity in all directions and in general is not
connected due to the presence of the forbidden region, which can
consist of two or more separate ``holes" in $\Sigma$ (see 
Ref.~\cite{GunzigetalCQG} for examples).

At best, the argument of Ref.~\cite{GunzigetalMPLA} arguing 
against
the presence of chaos for the special case of nonminimally coupled
fields can be applied (and generalized to arbitrary 
scalar-tensor cosmologies described by 
eqs.~(\ref{7})-(\ref{9})) to compact regions of the phase space 
$\Sigma$ as follows. Consider a compact region ${\cal C}$ in one 
of the sheets $\Sigma^+$ or $\Sigma^-$,
lying away from the boundary ${\cal B}$  of the forbidden 
region. Let $ \left( u, v \right) $ be smooth local coordinates 
covering ${\cal C}$ (if $u$ and $v$ do
not cover the entire region ${\cal C}$ one can consider an atlas
composed of two or more charts);
\begin{eqnarray}
   u & =&  u \left( H,\phi \right) \;,  \\
&&\nonumber\\
   v & = &  v \left( H ,\phi \right) \;,
\end{eqnarray}
is a smooth map from a region of ${\mathbb R}^2$ to the sheet
$\Sigma^\pm$, which is locally flat. Then, using the comoving 
time $t$ as a parameter,
\begin{eqnarray}
\dot{u} &= & \frac{\partial u}{\partial H} \dot{H} +
\frac{\partial u}{\partial \phi}\, \dot{\phi} =
\left[-\frac{\omega}{2\phi^2}+ \frac{\omega '}{2\phi
\left(2\omega+3\right)}\right] \dot{\phi}^2 \,\frac{\partial
u}{\partial H} 
+ \left( \frac{2H}{\phi}\frac{\partial u}{\partial H}
+ \frac{\partial u}{\partial \phi} \right) \dot{\phi} \nonumber 
\\
&&\nonumber \\ 
&& +\frac{\phi V'-2V}{2\phi \left( 2\omega+3 \right)}\, 
\frac{\partial u}{\partial H} \;, 
\label{17}\\  && \nonumber \\
\dot{v} &= &\frac{\partial v}{\partial H}\, \dot{H} +
\frac{\partial v}{\partial \phi} \, \dot{\phi} =
\left[-\frac{\omega}{2\phi^2}+ \frac{\omega '}{2\phi
\left(2\omega+3\right)}\right] \dot{\phi}^2 \frac{\partial
v}{\partial H} + \left( \frac{2H}{\phi}\frac{\partial v}{\partial H}
+ \frac{\partial v}{\partial \phi} \right) \dot{\phi} \nonumber 
\\
&&\nonumber \\
&& +\frac{\phi V'-2V}{2\phi \left( 2\omega+3 \right)}\, 
\frac{\partial v}{\partial H}
\;. \label{18} 
\end{eqnarray}
By using the expression~(\ref{11}) of $\dot{\phi}$ and the fact 
that $u$, $v$, and their derivatives only depend on $H$ and 
$\phi$, one can write eqs.~(\ref{17}) and (\ref{18}) in the form 
of the autonomous system of first order ordinary differential 
equations  for $ \left( u(t),v(t) \right)$
\begin{eqnarray}
 \dot{u} &= & f \left( u,v \right)  \;, \label{19}\\
&& \nonumber \\
\dot{v} & = & g \left( u,v \right) \;, \label{20}
\end{eqnarray}
where
\begin{eqnarray}
  f \left(u,v \right) &=& \left[-\frac{\omega}{2\phi^2}+
\frac{\omega '}{2\phi \left(2\omega+3\right)}\right] \left[
\frac{3H\phi \pm \sqrt{3H^2\phi^2\left(2\omega+3\right) - \omega
\phi V}}{\omega}\right]^2\frac{\partial u}{\partial H} \nonumber\\
&&\nonumber \\
&+& \frac{3H\phi \pm  \sqrt{3H^2\phi^2 
\left( 2\omega+3\right)-\omega\phi V}} {\omega} 
\left( \frac{2H}{\phi}\frac{\partial u}{\partial H} +
\frac{\partial u}{\partial \phi}\right) \nonumber \\
&&\nonumber \\
&+& \frac{\phi V'-2V}{2\phi \left( 2\omega+3 \right)}\, 
\frac{\partial u}{\partial H}  \;, 
\label{21}
\end{eqnarray}
and
\begin{eqnarray}
  g \left( u,v \right) & = & \left[-\frac{\omega}{2\phi^2}+
\frac{\omega '}{2\phi \left( 2\omega+3\right)}\right] \left[
\frac{3H\phi \pm \sqrt{3H^2\phi^2\left(2\omega+3\right) - \omega
\phi V}}{\omega}\right]^2\frac{\partial v}{\partial H} \nonumber\\
&&\nonumber \\
&+& \frac{3H\phi \pm \sqrt{ 
3H^2\phi^2\left(2\omega+3\right)-\omega\phi
V}} {\omega} \left( \frac{2H}{\phi}\frac{\partial v}{\partial H} 
+ \frac{\partial v}{\partial \phi}\right)  \nonumber \\
&&\\
&+& \frac{\phi V'-2V}{2\phi \left( 2\omega+3 \right)} \, 
\frac{\partial v}{\partial H} \;.
\label{22}
\end{eqnarray}
The reduction of the dynamical system to a first order autonomous
system with phase space consisting of a plane allows one to apply
the standard Poincar\'{e}-Bendixson theory which guarantees that
compact and connected regions of $\Sigma^\pm$ (corresponding to
compact and connected regions of ${\mathbb R}^2$) are free of
chaos. However, this argument only proves the statement of 
Ref.~\cite{GunzigetalMPLA} for such compact regions but not for the
entire phase space $\Sigma$. This larger phase space can be 
studied in the context of a more comprehensive analysis of the 
dynamics that we propose in the next two sections.


\section{Conformal mapping of the phase space}
\setcounter{equation}{0}

It is well known that the Jordan frame action (\ref{4}) of
scalar-tensor gravity can be mapped to the Einstein Hilbert action
by means of the conformal transformation
\begin{equation}
\label{23} g_{ab} \longrightarrow \tilde{g}_{ab} = \Omega^2
g_{ab},\hspace{10 mm} \;\;\;\; \Omega = \sqrt{\phi} \;,
\end{equation}
and the scalar field redefinition $\phi \longrightarrow
\tilde{\phi}(\phi)$ given by

\begin{equation} \label{24} 
d\tilde{\phi} = \sqrt{\frac{2\omega(\phi)+3}{16\pi }}\,\,
\frac{d\phi}{\phi}
\end{equation}
(see Refs.~\cite{MagnanoSokolowski,myreview,mybook} for 
reviews). In terms of conformally rescaled quantities in the 
Einstein frame, which are denoted by a tilde, the action 
(\ref{4}) assumes the form
\begin{equation}
\label{25} S= \int d^4x\sqrt{-g} \left[ \frac{\tilde{R}}{16\pi
G}-\frac{1}{2} \, \tilde{g}^{ab} 
\, \tilde{\nabla}_a\tilde{\phi} \, \tilde{\nabla}_b\tilde{\phi}
- U \left( \tilde{\phi} \right) \right] \;,
\end{equation}
where
\begin{equation}
\label{26} U(\tilde{\phi}) =
\frac{V\left[\phi \left( \tilde{\phi} \right) \right]}{ \phi^2} 
\;.
\end{equation}
The sign of the potential is preserved by the conformal
transformation. By studying the dynamics in the Einstein frame and
mapping back to the Jordan frame, one can study the dynamics and, in
particular, the absence or presence of chaos.

Before we proceed, let us study the conformal cousins in the
Einstein frame of the de Sitter fixed points living in the Jordan
frame. It is easily concluded that de Sitter equilibrium points in
the Jordan frame are mapped into de Sitter equilibrium points in 
the Einstein frame, and vice-versa. In fact, by using 
eq.~(\ref{24}) and the usual range of values of the coupling 
function
$2\omega(\phi)+3 > 0$ given by Solar System experiments and
guaranteeing a positive-definite kinetic energy term for the 
scalar, one has $ d\tilde{\phi}/d\tilde{t} =0$ if and only if
$d\phi/dt =0$. In addition, eqs.~(\ref{1}) and
(\ref{23}) yield
\begin{equation}
\label{27}
 d\tilde{s}^2 = \Omega^2ds^2 = \Omega^2\left(-dt^2 +
a^2d\vec{x}^2\right) = -d\tilde{t}^2 +
\tilde{a}^2d\vec{x}^2 \;,
\end{equation}
where $ d\tilde{t} \equiv \sqrt{\phi}\, dt $ and $\tilde{a} 
\equiv \sqrt{\phi} \, a$. Hence, 
\begin{equation}
\label{28} \frac{d\tilde{\phi}}{d\tilde{t}} = \sqrt{\frac{2\omega
+3}{16\pi }} \, \phi^{-3/2 } \, \frac{d\phi}{dt} \;.
\end{equation}
Because we consider 
positive gravitational coupling $\phi>0$, the sign of
$ d\tilde{\phi}/d\tilde{t} $ is the same as that of
$\dot{\phi}$, and $ d\phi/dt = 0$ if and only if
$ d\tilde{\phi}/ d\tilde{t} =0$, while
\begin{equation}
\label{29} 
\frac{d\tilde{H}}{d\tilde{t}} =
\frac{1}{\phi}\left(\frac{\ddot{\phi}}{2\phi} -
\frac{3\dot{\phi}^2}{4\phi^{3/2} }+ \dot{H}
-\frac{H\dot{\phi}}{2\phi}\right) \;.
\end{equation}
Therefore, a de Sitter fixed point with constant scalar field
$\left( \dot{H},\dot{\phi} \right) = \left( 0,0 \right)$ in the 
Jordan frame corresponds to a
de Sitter fixed point with constant scalar field
$ \left( d\tilde{H}/d\tilde{t},
d\tilde{\phi}/d\tilde{t}\right) = \left( 0,0 \right) $ in the 
Einstein frame (this fact is related to the scale invariance  
property of the exponential function). It is not true, 
as is instead stated in Ref.~\cite{Walliser} that the conformal 
transformation becomes singular at the equilibrium points.

The condition~(\ref{14ter}) for the existence of the Jordan 
frame fixed point $\left(H_0,\phi_0 \right) $ translates into 
the corresponding condition $ dU/d\tilde{\phi}=0 $. In 
fact,
\begin{equation}
\frac{dU}{d\tilde{\phi}} =
\frac{dU}{d\phi} \, \frac{d\phi}{d\tilde{\phi}}=
\sqrt{\frac{16\pi }{2\omega(\phi)+3}} \, \frac{1}{\phi}
\, \left[V'(\phi)-\frac{2V(\phi)}{\phi}\right]
\end{equation}
and eq.~(\ref{14ter}) implies $ 
\left. \frac{dU}{d\tilde{\phi}}\right|_{\tilde{\phi}_0} =0$,
whereas $\phi=\phi_0=$~constant is equivalent to 
$\tilde{\phi}=\tilde{\phi}_0=$~constant. Moreover, stability of 
the de Sitter 
fixed point in the Jordan 
frame corresponds to stability of the corresponding de Sitter 
point in the Einstein frame, as shown in the following. Since 
\be
\tilde{H}\equiv \frac{1}{\tilde{a}}\, 
\frac{d\tilde{a}}{d\tilde{t}} 
= \frac{\dot{\phi}}{ 2\phi^{3/2} }+ 
\frac{H}{\sqrt{\phi}} \;,
\ee
perturbations $\delta H $ and $ \delta \phi$ in the Jordan frame
correspond to the Einstein frame perturbations
\begin{eqnarray}
\delta \tilde{H} &= &  
\frac{\delta \dot{\phi}}{2\phi^{3/2 }}
+ \frac{\delta H}{\sqrt{\phi_0}} -
\frac{H_0}{2\left.\phi_0\right.^{3/2}} \,\delta \phi \;,
\label{30} \\
&&\nonumber \\
\delta \tilde{\phi} & = &  
\sqrt{\frac{2\omega(\phi_0)+3}{16\pi }}\, 
\frac{\delta  \phi}{\phi_0} \;,  \label{31}
\end{eqnarray}
to first order. If the Jordan frame fixed point $ 
\left (H_0,\phi_0 \right) $ is
stable, perturbations $\delta H $ and  $\delta \phi$ 
do not grow (or decay rapidly) and the same can be concluded for 
the Einstein frame perturbations
$\delta \tilde{H}, \delta \tilde{\phi}$; and vice-versa.


\section{Dynamics in the conformally rescaled world}
\setcounter{equation}{0}

We finally proceed to study the dynamics in
general relativity in the conformally rescaled world. For economy of
notations we drop the tilde and in this section an overdot 
denotes differentiation with respect to the Einstein frame 
comoving time. The Einstein-Friedmann equations are
\begin{eqnarray}
&& H^2 = \frac{\kappa}{6}\left[\dot{\phi}^2 + 2U(\phi)\right] 
\;, \label{32}  \\
&&\nonumber \\
 &&\dot{H} = -\frac{\kappa}{2} \, \dot{\phi}^2 \;, \label{33}\\
&& \nonumber \\
 &&\ddot{\phi} + 3H\dot{\phi} + 
\frac{dU}{d\phi} = 0 \;.  \label{34}
\end{eqnarray}
Since general relativity is a trivial case of 
scalar-tensor gravity, the structure of the phase space 
discussed in section 2 is
still present, albeit some simplifications occur. The Hamiltonian
constraint (\ref{32}) can be seen as an algebraic equation for 
$\dot{\phi}$
with roots
\begin{equation} \label{35} 
\dot{\phi}_{\pm}(H,\phi) = \pm
\sqrt{2\left[\frac{3H^2}{\kappa}-U(\phi)\right]} \;,
\end{equation}
which makes evident the double-sheeted structure of the
two-dimensional phase space $\Sigma = \Sigma^+ \cup \Sigma^-$
embedded in the three-dimensional space $(H,\phi,\dot{\phi})$. In
general there is a forbidden region
\begin{equation}
{\cal F} \equiv \{\left( H,\phi \right): 
\;\;\;\frac{3H^2}{\kappa} 
< U(\phi)\}
\end{equation}
with boundary
\begin{equation} \label{36}
{\cal B} \equiv \Sigma^+ \cap \Sigma^- = 
\{\left( H,\phi,\dot{\phi} \right): \;\;\;
\frac{3H^2}{\kappa} = U(\phi) \;, \; \; \;\dot{\phi}=0\} \;.
\end{equation}
The de Sitter equilibrium points $ \left( H_0,\phi_0 \right) $ 
are the only de Sitter spaces, while in general scalar-tensor 
cosmology de Sitter spaces with non-constant 
scalar field may be solutions, although they are
not fixed points (cf. eq.~(\ref{8})). Moreover, contrary to the 
case of section~2, 
all 
the de Sitter fixed points are forced to lie on the boundary 
${\cal B} = \Sigma^+ \cap 
 \Sigma^-$ between the upper and lower 
sheet, corresponding to $\dot{\phi}=0$. 
Eqs.~(\ref{32})-(\ref{34}) yield the two conditions for the 
existence of de Sitter spaces $\left( H_0,\phi_0 \right) $ 
(there are only two  conditions because only two of the 
eqs.~(\ref{32})-(\ref{34}) are  independent),
\begin{eqnarray} 
&& H_0^2 = \frac{\kappa}{3} \,  U_0 \;, \label{37}  \\
&&\nonumber \\
&& U_0'= 0 \;,  \label{38} 
\end{eqnarray}
where $U_0 \equiv U \left( \phi_0 \right) $ and $U'_0 \equiv
\frac{dU}{d\phi}|_{\phi_0}$. de Sitter fixed points exist if the
potential $U(\phi)$ has maxima, minima, or inflection points at
$\phi_1, \phi_2, \, \ldots \,, \phi_n, \, \ldots \,$ with 
$U'(\phi_1)=U'(\phi_2)= \ldots =U'(\phi_n)=\ldots=0 $. One 
expects that if $U$ has a
maximum (respectively, a minimum) in $\phi_i$, the 
corresponding fixed point 
$ \left( \sqrt{ \kappa U \left( \phi_i  \right) / 3 }, \phi_i 
\right)$ will be
unstable (respectively, stable) as we prove later.

Eq.~(\ref{33}) allows one to conclude immediately that there
are no limit cycles (periodic orbits) because $H$ is always 
decreasing (apart from the fixed point solutions) and can not 
come back to a previous value during the evolution of the 
system. Each point of ${\cal B}$ where $U'\neq 0$ can be crossed 
by a trajectory only once, and in a definite direction specified 
by the sign of $U'$ at that point.

In the upper sheet with boundary removed $\Sigma^+ \backslash 
{\cal B}$, it is $\dot{\phi}>0$ and $\dot{H} <0$, while in 
$\Sigma^- \backslash {\cal B}$ it is $\dot{\phi}<0$ and $\dot{H} 
<0$, hence $\dot{\phi}$ can not change sign away from the 
boundary ${\cal B}$. On ${\cal B}= \Sigma^+ \cap \Sigma^-$ it is
$\dot{H}=0,\dot{\phi}=0,\ddot{\phi}=-U(\phi)$, as follows from 
the
dynamical equations. The tangent to the orbits of the solutions
parametrized by the time $t$ in the space $ 
\left( H,\phi,\dot{\phi} \right) $ is the vector
\begin{equation} \label{39} 
\vec{T}(t) = \left( \dot{H}(t), \dot{\phi}(t), 
\ddot{\phi}(t)\right)
\end{equation}
and, on ${\cal B}$,
\begin{equation}
\vec{T}(t)|_{\cal B} = \left( 0,0,-U' \right) \;.
\end{equation}
On the boundary ${\cal B}$, the tangent $ \vec{T}$ can only be 
vertical and 
no motion along the $H$ or $\phi$ directions can occur. This 
means that there can be no motion along the curve ${\cal B}$ and 
portions of ${\cal B}$
can not be parts of orbits of solutions. At points of ${\cal B}$ 
where 
$U'=0$, it is $\left|\left| \vec{T} \right|\right| =0$ and there 
is no motion: these points are de Sitter fixed points. At
points of ${\cal B}$ where $U'> 0$, the tangent $\vec{T}$ points
downward along the negative $\dot{\phi}$-axis and an orbit can 
only go from the upper sheet $\Sigma^+$ to the lower sheet 
$\Sigma^-$ by crossing the boundary ${\cal B}$. If instead 
$U'<0$ at points of ${\cal B}$, the tangent $\vec{T}$ to 
the orbit is pointing upward in the positive 
$\dot{\phi}$-direction and the orbit crosses
from $\Sigma^-$ to $\Sigma^+$. This excludes the possibility that an
orbit ``bounces" on the boundary ${\cal B}$  back to the 
sheet it came from. The
possibility is not excluded that, in certain potentials, once 
the orbit has crossed, it changes component
of the vertical velocity $\ddot{\phi}$ and comes back to the
boundary to change sheet again. From eqs.~(\ref{32})-(\ref{34}) 
it follows that
\begin{equation}
\label{40} 
\dot{H} + 3H^2 = \kappa \, U \;.
\end{equation}

Since $\dot{H}$ always decreases monotonically (except, of 
course, at the fixed points) either $H$ tends to a finite 
limit (horizontal asymptote) $H_0$ as $t\rightarrow +\infty$, or 
else $H \rightarrow -\infty$. (For a given potential $U(\phi)$, 
there may be orbits
going to a finite $H_0$ and other orbits going to $-\infty$,
depending on the initial conditions).

If $H\rightarrow H_0$ as $t\rightarrow +\infty$, then $\dot{H}
\rightarrow 0$ and eq.~(\ref{33}) implies that 
$\dot{\phi}\rightarrow 0 $, which implies that also the scalar 
field $\phi$
approaches a horizontal asymptote $\phi_0$. In this situation, 
eq.~(\ref{40}) yields $3H_0^2 = \kappa
U \left( \phi_0 \right) $, i.e., we have a de Sitter fixed point 
satisfying eq.~(\ref{37}), which can only exist if 
$U'(\phi_0) =0$. If
$U'\not=0$ for all values of $\phi$ (i.e., if $U(\phi)$ is strictly
monotonic), there is the possibility that $H\rightarrow
-\infty$ instead, which corresponds to a Big Crunch in a finite
time, but there is also the possibility that $H(t) \rightarrow 
H_0$ asymptotically while $\phi (t) \rightarrow \pm\infty $ and 
both $\dot{H},\dot{\phi}
\rightarrow 0$ as $t \rightarrow +\infty$. An example is the 
exponential potential
\begin{equation}
U \left( \phi \right) = A 
\exp\left[ \pm\sqrt{\frac{16\pi}{p}} \, \phi\right] 
\;\;\;\;\;\;\;\; \left ( p>1 \right),
\end{equation}
which gives power-law inflation \cite{LiddleLyth}
\begin{eqnarray}
 a ( t ) & = & a_0 \, t^p  \;, \\
&&\nonumber \\
\phi(t) &= &  \sqrt{\frac{p}{4\pi}} \, \ln\left(\sqrt{\frac{8\pi
  A}{p\left(3p-1\right)}} \,\, t \right) \; ,
\end{eqnarray}
for which $H= p/t \rightarrow 0 $ while
$\phi\rightarrow+\infty$ and $ \dot{\phi} \rightarrow 0 $.

We can now consider the situation in which 
$ H\rightarrow -\infty $ and consider times sufficiently large 
so that $H<0$. To proceed we need to make some 
assumptions on the potential: we first
assume that $U$ is {\em monotonic}, so $U'$ has definite sign. 
The 
first possibility is $U'<0$ for all $\phi$; then the 
Klein-Gordon equation $ \ddot{\phi}= -3H\dot{\phi}-U' $ implies 
that in the upper sheet $\Sigma^+$, where $\dot{\phi}>0$, it is 
$\ddot{\phi}> 0 $ and therefore $\phi\rightarrow +\infty$; the 
point $ \left( H,\phi \right)$ representing the universe in the 
phase space  goes to infinity without oscillating. In the lower 
sheet $\Sigma^-$, where
$\dot{\phi}<0$, either $\phi$ has a horizontal asymptote with
$\phi\rightarrow\phi_0$, or else $\phi\rightarrow -\infty$. The
first possibility is excluded because it would imply that
$\dot{\phi}\rightarrow 0$ and, according to eq.~(\ref{33}),
also $\dot{H} \rightarrow 0$, which contradicts the assumption 
that $H\rightarrow -\infty $; hence also in this case $(H,\phi)$ 
goes to
infinity without oscillations.

In the case $U'>0$ for all values of $\phi$, take points in the 
upper sheet $\Sigma^{+}$ with $\dot{\phi}>0$:  then either 
$\phi(t) \rightarrow \phi_0$ (horizontal asymptote) or else 
$\phi(t)\rightarrow +\infty$. If $\phi$ has a horizontal 
asymptote $ \phi_0 $, i.e., if $ \phi\rightarrow \phi_0$, then 
$\dot{\phi}(t)\rightarrow 0$ and also $\dot{H}(t) \rightarrow 
0$: but this contradicts the assumption that $H(t)\rightarrow 
-\infty$. Then it must be $\phi(t)\rightarrow +\infty$ and the 
point $\left( H,\phi \right) $ goes to infinity without 
oscillations. If instead we pick a point in the lower sheet 
$\Sigma^{-}$ with $\dot{\phi}<0$, then $\ddot{\phi}=-3H 
\dot{\phi} -U'<0 $ and $\phi (t) $ goes to minus infinity 
without oscillations.

We can then consider the trivial case in which $U'=0$ for all 
values of $\phi$: this represent a cosmological constant and 
gives unviable early universe models in which inflation does not 
stop or dark energy models that are disfavoured (a scalar field 
is introduced to get away from the cosmological constant and 
its problems \cite{Weinberg2}); 
therefore, this is a case of mathematical
interest that we include for completeness. In this case 
$\ddot{\phi}=-3H\dot{\phi}$, which integrates to
$\dot{\phi}= C/a^3 $ where $C$ is a non-zero constant, and
$\dot{\phi}\rightarrow \infty$: the point $(H,\phi)$ tends to
infinity without oscillations.

Generally, we can consider a non-monotonic potential with a minimum,
or possibly several maxima and minima. In this case the sign of
$U(\phi)$ changes as $\phi$ evolves; this situation is best 
studied with a Ljapunov function.

First, assume that $U(\phi)$ has a single absolute minimum
$ U_0\equiv U (\phi_0)$ attained at the single value $\phi_0$ of 
the
scalar field (examples are the widely used potentials $U =
m^2 \phi^2 / 2 $ or $ U=\lambda \phi^4$), then $ \left( H_0,
\phi_0 \right) = \left( \sqrt{\kappa U_0/3}, \phi_0\right)$ is a 
fixed point. The function
\begin{equation}
L \left( H,\phi \right) = \frac{ \dot{\phi}^2}{2} +U (\phi)-U_0
\end{equation}
is a Ljapunov function. In fact, $ U (\phi ) > 0 \; \; \forall 
\phi  \not= \phi_0$ and $L \left( H,\phi \right)>0 \;\; 
\forall \left( H,\phi \right) \not= \left( H_0,\phi_0 \right)$.
Moreover, $L(H_0,\phi_0)=0$ since at the fixed point 
$\dot{\phi}=0$.
Along the orbits of the solutions we obtain, upon use of the
Klein-Gordon equation,
\begin{equation}
\frac{dL}{dt} = \dot{\phi} \left(\ddot{\phi} + U'\right) =
-3H\dot{\phi}^2 <0
\end{equation}
when $H>0$, i.e., for all expanding universes. This guarantees 
that $ \left( H_0,\phi_0 \right) $ is an attractor with an 
attraction  basin at least as
wide as the $ H > 0 $ region of $\Sigma$.

The Ljapunov function $L(H,\phi)$ also allows us to immediately
conclude that contracting de Sitter spaces with $H_0 =
-\sqrt{ \kappa U_0/3} $ are unstable because $H<0$ and
$\dot{L}>0$ in a neighbourhood of this equilibrium point (it is 
well known that contracting de Sitter spaces are unstable --- 
see, e.g., Ref.~\cite{deSitter}).

If $U(\phi)$ has an absolute minimum that is assumed at two (or 
more) different values of $\phi$, say $\phi_1, \phi_2, \, \ldots 
\, , \phi_n, \, \ldots \, $
then $U(\phi)> U(\phi_1) = U(\phi_2) = \ldots = U(\phi_n) =\ldots
\equiv U_{min} \;\;\forall \phi \not\in
\{\phi_1,\phi_2,\, \ldots \, ,\phi_n, \, \ldots\,  \}$ and 
$U(\phi)$ must assume
maxima between the minima. In this case the phase space $\Sigma$
will contain the attraction basins of the stable fixed points
corresponding to the minima of $U(\phi)$ (the maxima corresponding
to unstable fixed points), with a separatrix going through a 
fixed point between two adjacent attraction basins 
\cite{Walliser}.


\section{Discussion and conclusions}
\setcounter{equation}{0}

Based on the reported marginal observational evidence for 
present superacceleration of the universe, which can not be 
explained by general relativity with a canonical scalar field, 
we consider spatially flat homogeneous and isotropic cosmologies 
in scalar-tensor gravity. The dynamics are conveniently 
studied by a conformal mapping of the Jordan frame 
scalar-tensor theory into  the Einstein frame, in which gravity 
reduces to general  relativity and the rescaled scalar field is 
minimally coupled to gravity (a non-minimal coupling of the 
scalar to ordinary matter is irrelevant here because we consider 
the scalar to be the dominant form of ``matter'' in the 
universe, 
and other material sources can be neglected). Apart from the 
original motivation to study scalar-tensor gravity by using the 
mathematical trick of the conformal transformation to the 
Einstein frame, the study of the phase space and of the dynamics 
of general relativity with a minimally coupled scalar field is 
interesting in itself. We emphasize that this work is limited to 
spatially flat ($K=0$) FLRW models because they describe  
our universe according to the recent observations of the cosmic 
microwave background. However, from the dynamical point of view, 
the $K=+ 1$ cases are even more interesting \cite{chaos}, and 
even spatially flat universes containing Yank-Mills fields 
exhibit chaotic oscillations of these fields \cite{YM}.

The results obtained about the Einstein frame dynamics can be 
mapped back to the Jordan frame. The 
general picture obtained in the  Einstein
frame is that of a phase space that is free of chaos, with the
orbits of the solutions converging to attractor points or going to
infinity. Depending on the form of the scalar field potential, there
can also be power-law attractor solutions, which are well known 
to exist in several scalar-tensor gravity theories 
\cite{mybook}. When
mapped back to the Jordan frame, the non-chaotic Einstein 
frame dynamics translates into non-chaotic Jordan frame 
dynamics; when they exist, stable (unstable) 
equilibrium points are mapped
into stable (unstable)  equilibrium points, their attraction
basins are deformed but they are still present, and the
boundaries between different attractor points are 
well defined
separatrices also in the Jordan frame. The possibility of such
boundaries having fractal dimension, which would be a clear
signature of chaos \cite{Glendinning}, is ruled out (note that 
fractal basin boundaries have instead been found in $K\neq 0$ 
FLRW, or in anisotropic universes \cite{fractalboundaries}).

The dynamics of spatially flat scalar-tensor cosmologies is 
thus well defined and chaos-free, due to the dimensional 
reduction of the phase space $\Sigma$ to two dimensions, as 
conjectured in  Refs.~\cite{GunzigetalMPLA,GunzigetalCQG}. 
However, the proof of this
conjecture is non-trivial due to the complicated structure of 
the phase space $\Sigma = \Sigma^+ \cup\Sigma^-$. Only spatially 
flat cosmologies enjoy the reduction of the phase space to two
dimensions: the scale factor $a(t)$ of the FLRW metric~(\ref{1}) 
only appears in the combination $H\equiv \dot{a}/a$ and 
its first derivative $\dot{H}$ in the field equations with 
$ K=0$. If $K \not= 0$ instead, terms of the form 
$ K/a^2 $ will appear, spoiling the reduction of the phase space 
to two dimensions.
The surface $\Sigma$ in the three-dimensional space
$\left( H,\phi,\dot{\phi} \right)  $ separates the orbits of the 
solutions corresponding to $ K=+1$ (located above $\Sigma^+$ or 
below $\Sigma^-$) from orbits corresponding to $ K = -1$ 
(located between $\Sigma^+$ and $\Sigma^-$). This property was 
first realized in Ref.~\cite{Belinskyetal} for the special case 
of  chaotic inflation with a massive scalar field in general
relativity, and it corresponds to the impossibility of having 
dynamical transitions between different topologies of the 
spatial sections of a FLRW universe. Orbits 
of the solutions
corresponding to $ K =\pm 1$ are free to move in the three
dimensions $(H,\phi,\dot{\phi})$, where there is ``enough room" 
for them to wind around each other, so chaos can occur, and it 
has indeed been reported in the literature \cite{chaos}.

Because inflation takes the early universe extremely close to, but
not exactly to, a $ K=0$ FLRW model, the orbits of the solutions
can in principle depart slightly from the surface $\Sigma$, but 
they are extremely close to it. Then, the 
possibility of chaotic
dynamics is not ruled out exactly, but is postponed for an extremely
long time. Of course, if a second scalar field is present and
significantly contributes to the dynamics of the universe, then the
dimension of the phase space jumps up by two and chaos becomes
possible, and is also reported in the literature (e.g.,
Ref.~\cite{EastherMaeda}) --- chaos in the dynamics of two 
mutually 
coupled scalar fields is an important element of reheating 
after inflation.

To summarize, in addition to taking a standpoint on the 
possibility of chaos and
generalizing the context of this debate to any $ K =0$
scalar-tensor cosmology described by the action (\ref{4}), we 
have provided a general picture 
of the
phase space and of the dynamics (see also 
Refs.~\cite{KolitchEardley}-\cite{HoldenWands} 
for the special case of  Brans-Dicke theory, the prototype of 
scalar-tensor gravity). Special choices of the arbitrary 
functions $\omega (\phi) $ and $V( \phi) $ will be the subject 
of future research.


\section*{Acknowledgments}

We thank an anonymous referee for pointing out various 
references. This work was supported by the Natural Sciences  and
Engineering Research  Council of Canada ({\em NSERC}).


\end{document}